# Separating enhancement from loss:

# plasmonic nanocavities in the weak coupling regime

Kasey J. Russell*, Tsung-Li Liu, Shanying Cui and Evelyn L. Hu

Harvard University School of Engineering and Applied Sciences

9 Oxford St., McKay Laboratory, Cambridge MA 02138, USA

*E-mail: krussell@seas.harvard.edu

By modifying the density of optical states at the location of an emitter, weak cavity-emitter coupling can enable a host of potential applications in quantum optics, from the development of low-threshold lasers to brighter single-photon sources for quantum cryptography[1]. Although some of the first demonstrations of spontaneous emission modification occurred in metallic structures[2,3], it was only after the recent demonstration of cavity quantum electrodynamics effects in dielectric optical cavities[4] that metal-based optical cavities were considered for quantum optics applications[5-13]. Advantages of metal-optical cavities include their compatibility with a large variety of emitters and their broadband cavity spectra, which enable enhancement of spectrally-broad emitters. Here, we demonstrate a metal-based nanocavity structure that achieves radiative emission rate enhancements of 1000, opening up the possibility of pursuing cavity electrodynamics investigations with intrinsically broad optical emitters, including organic dyes and colloidal quantum dots.

In dielectric optical cavities, the signature of weak cavity-emitter coupling is unambiguous: changes in radiative emission rate ($\nu_R$) are correlated with proportional changes in emission intensity. Metal structures, however, are inherently lossy, and can cause high rates of non-radiative recombination ($\nu_{NR}$). In addition, metallic structures can act as optical antennas to alter the efficiency of excitation and collection[14-16]. An ambiguous situation can therefore arise in which antenna effects increase the emission intensity while loss increases the total decay rate $\nu_F = \nu_R + \nu_{NR}$.

The experiments described here prove that our cavities are operating in the weak coupling regime. The spontaneous emission rate enhancement we observe is significantly larger than what has been seen in dielectric cavities[17], but it is comparable to the field enhancements reported in surface-enhanced Raman scattering from similar structures[18]. Our cavity design (Fig. 1) is based on surface plasmon coupling between a Ag substrate and a Ag nanowire lying parallel to the substrate[19]. The spacing $d_G$ between the nanowire and substrate is established by coating the substrate in a thin, uniform layer of $SiO_2$ and fluorescent organic dye [tris-(8-hydroxyquinoline) aluminium, $Alq_3$]. This design ensures that the dye will be located at the high-field regions of the cavity modes (Fig. 1b). Similar structures have been shown to confine gap-mode plasmons[11,19], but their ability to yield cavity quantum electrodynamics effects has not previously been reported.

Fluorescence spectra from individual cavities show a strong modification of the $Alq_3$ spectrum as compared to both $Alq_3$ on glass and to an off-cavity area of $Alq_3/SiO_2/Ag$ (Fig. 2a; for illustration of off-cavity geometry, see Fig. 1c). In similar structures, the wavelengths of the peaks in the cavity spectra have been shown to correspond exactly with the cavity resonances[19], establishing that there is sufficient dye-cavity coupling to modify the spectral properties of the emission.

Fluorescence lifetime measurements on cavities also show a strong modification due to the cavity, with an enhancement of both decay rate $v_F$ and emission intensity (Figs. 2B and S1 and Supplementary Section 1). The shortest-lifetime component of the traces is only observed when measuring on a cavity and is therefore referred to as the cavity lifetime ($\tau_C$). On-cavity data also contain longer-lifetime components that have decay rates matching those of measurements of off-cavity $Alq_3$ on $SiO_2/Ag$. This is "background" emission from dye surrounding the cavity (Fig. 1c). Off-cavity emission decays more rapidly than fluorescence from $Alq_3$ on glass because a short-lifetime decay pathway, such as surface plasmon emission or electron-hole pair creation in the metal, leads to an increase in $v_F$, and this decay pathway does not result in far-field radiation. The peak intensity appears ~1.5 times higher than that of $Alq_3$/glass because while the Ag in the $SiO_2/Ag$ structure reflects the fluorescence up toward the collection optics, ~60 % of the fluorescence from $Alq_3$/glass is emitted through the glass substrate and is not collected (Supplementary Section 2). Accounting for this difference in collection efficiency, we see that the off-cavity fluorescence is actually dimmer than fluorescence from $Alq_3$/glass, allowing us to unambiguously identify loss as the cause of the increase in $\nu_F$ as compared to dye on glass. This is the characteristic behavior of fluorescence quenching near a metal surface[2].

To conclusively show that our cavities enable emission enhancement, we measured three cavity structures of different gap spacing $d_G$ (samples A, B, and C, with $d_G \sim$ 5 nm, 15 nm, and 25 nm, respectively). Since the field confinement within the cavity is a strong function of $d_G$, this series of samples should exhibit large differences in cavity enhancement[8,9]. Antenna effects, meanwhile, should be similar in the samples (Fig. S2 and Supplementary Section 3). These samples therefore enable us to separately probe cavity and antenna effects.

The quantum efficiency of cavity emission ($\Phi_C$) is proportional to the number of cavity-coupled photons emitted, which is proportional to the time-integrated intensity of the cavity lifetime component ($I_C$). In the case of recombination dominated by loss (where changes in $\tau_C$ are caused by changes in $\nu_{NR}$), we can write $\Phi_C = \tau_C \nu_R$, illustrating that $I_C$ and $\tau_C$ are proportional in this regime. In the opposite limit of ideal cavity-enhanced recombination, changes in $\tau_C$ are driven by changes in $\nu_R$, and (since $\tau_C = (\nu_R + \nu_{NR})^{-1}$) we can instead write $\Phi_C = 1 - \tau_C \nu_{NR}$. Assuming $\nu_{NR}$ remains constant, in this regime $I_C$ will actually increase for lower $\tau_C$. Thus we have two expressions to describe the behavior of cavities in the complementary cases of loss-dominated and cavity-enhanced emission, respectively.

The cavities with lowest $I_C$ exhibit loss-dominated dependence, which is shown as a solid line in Fig. 3a. A fraction of all cavities measured (~8 of 21 from sample A, ~4 of 22 from sample B, and ~4 of 12 from sample C) fall along that line. However, the majority of cavities do not follow this trend, and therefore loss cannot account for the dominant observed behavior.

The opposite limit of ideal cavity-enhanced recombination is indicated with a broken line in Fig. 3a for the parameters $I_C(\tau_C=0) = 74$ a.u. and $\tau_{NR} = \nu_{NR}^{-1} = 62$ ps. The cavities with highest $I_C$ fall along that line, indicating that the radiative lifetime $\tau_R$ in these cavities is less than 100 ps. Given that $\tau_R$ is estimated to be more than 100 ns in Alq$_3$[20], this requires a radiative lifetime reduction of more than 1000, which is slightly larger than, but consistent with, theoretical simulations of devices with larger $d_G$[9,21] and is consistent with data from cavities containing PbS colloidal quantum dots (Fig. S3 and Supplementary Section 4). The values of $I_C(\tau_C=0)$ and $\nu_{NR}$ used to generate the broken line were chosen to fit the data, and they also yield reasonable values for the maximum $\Phi_C$ observed (~ 0.35 at $\tau_C$ = 40 ps) and for $\Phi_C$ at the intersection of the solid and broken lines: ~ 0.045 at $\tau_C$ = 59 ps. The reported value of $\Phi_C$ for Alq$_3$ on glass (~0.1-0.2) lies between these two values[20]. Thus, although the non-radiative decay rate is increased near metals, within a metal cavity there may also be an increase in radiative emission rate, yielding a net increase in fluorescent efficiency. Similar effects have been seen in bow-tie antenna structures[22].

To guard against cavity-specific antenna effects within $I_C$, we take the ratio of $I_C$ to the time-integrated intensity of the background lifetime components ($I_B$) extracted from the same fluorescence lifetime trace (Fig. 3b). Because antenna effects should increase $I_B$ along with $I_C$ (Fig. S2 and Supplementary Section 3), the ratio $I_C/I_B$ provides an approximate measure of the enhancement of the cavity beyond that of the background. Maximum $I_C/I_B$ occurs for the same $\tau_C$ at which $I_C$ is maximal, as expected for cavity enhancement.

In reality, it is unlikely that either $\nu_{NR}$ or $\nu_R$ is constant: the same condition that enhances $\nu_R$—greater field confinement via smaller $d_G$—also increases $\nu_{NR}$. This is because reducing the cavity gap concentrates the field but also increases the fraction of energy contained in the metal[6]. One therefore expects a trade-off between cavity enhancement and metal loss, with both $\nu_R$ and $\nu_{NR}$ decreasing with gap spacing. This trade-off explains why the dependence of $I_C$ on $\tau_C$ is not monotonic (Fig. 3a). For a high-brightness photon source, the optimal $d_G$ would maximize the total intensity enhancement of the cavity.

To fully understand the operating mechanism of our cavities, we separately addressed cavity and antenna effects. However, it is the appropriate balance of the two that will yield the ultimate goal of a high-brightness photon source. The total enhancement of intensity can be found by dividing the peak intensity of the cavity lifetime component ($a_C$) by the peak intensity from the Alq$_3$/glass sample and normalizing by the number of active emitters (Fig. 3c). The total enhancement peak is ~ 500, which, coupled with the large reduction in emission lifetime, indicates that these cavities provide a promising platform for the development of on-demand photon sources for quantum optics applications.

Thus we have demonstrated clear evidence of individual metal-optical nanocavities operating in the weak-coupling regime. We have tracked the measured spontaneous emission enhancement through cavities with different gap spacing, allowing us to separately delineate the effects of loss to the metal as well as antenna effects. Because this cavity design is applicable to a wide variety of emitters, our results not only demonstrate large cavity quantum electrodynamics effects in a metal-optical structure; they also suggest a route to extend such effects to emitters that are incompatible either spectrally or structurally with conventional dielectric microcavity designs, including diamond nanocrystals, colloidal quantum dots, and organic dyes.

# Methods

## Sample fabrication

Four Alq$_3$ samples were fabricated: a control sample of Alq$_3$ deposited on a glass slide (Alq$_3$/glass) and three cavity structures of different gap spacing $d_G$ (samples A, B, and C, with $d_G \sim$ 5 nm, 15 nm, and 25 nm, respectively).

A silver substrate of sub-nm root-mean-square (rms) roughness (typically 0.5-0.8 nm rms over a 2x2 µm$^2$ area) was prepared via template-stripping from an atomically-flat Si wafer. The thickness of the Ag layer was ~300 nm, and it was anchored to a Si "handle" using epoxy (EPO-TEK 377, Epoxy Technologies) that was cured by baking the sample at 150C for ~3 hr. Removal from the template was accomplished by sliding the blade of a razor under the corner of the Si handle.

The freshly-exposed, flat Ag surface was immediately covered in SiO$_2$ by sputtering from a quartz target using an Ar-ion plasma at 4 mTorr process pressure. The SiO$_2$ thickness was varied between samples to yield the desired $d_G$ with fixed Alq$_3$ thickness.

The film of Alq$_3$ was thermally evaporated onto all samples simultaneously at a pressure of ~ 2x10$^{-6}$ Torr and rate of ~ 4.8 nm/min to a total thickness of 3 nm, as measured with an in-situ quartz crystal microbalance. After evaporation of Alq$_3$, the samples were stored in the dark in a vacuum dessicator when not being measured. Under these conditions, both the spectral and time-resolved fluorescence was found to be stable for at least several days.

Silver nanowires (Blue Nano) of diameter ~100 nm and length ~1 – 30 µm were deposited onto a piece of polydimethylsiloxane (PDMS) in a droplet of ethanol that was allowed to dry. The nanowires were then transferred to the Alq$_3$/SiO$_2$/Ag structure via stamping.

**Measurement procedure**

Light from the pulsed, frequency-doubled beam ($\lambda \sim 460$ nm) of a Ti:Sapphire laser was directed through a 100x, 0.9NA microscope objective onto the sample in the direction normal to the substrate (i.e. along the z-axis). Fluorescence from the sample was collected through the same microscope objective, spectrally filtered to remove the laser signal, and spatially filtered to collect light only from the excitation region (i.e. in a confocal microscopy arrangement) by coupling into a 25 μm-core multi-mode optical fiber. The signal was analyzed using either a grating spectrometer with liquid-nitrogen cooled CCD camera (Princeton Instruments) or a fast avalanche photodiode (APD) having an impulse response with a full width at half maximum of ~ 50 ps (Micro Photon Devices). Spectra were typically acquired by integrating for 10 s, whereas time-resolved measurements were spectrally integrated and acquired for 2-5 min using time-correlated single photon counting (TCSPC).

All measurements were performed in a sample chamber continuously purged with nitrogen gas to reduce degradation of the $Alq_3$[23]. Multiple subsequent measurements of fluorescence lifetime yielded the same results and were typically performed at a laser power of ~ 500 nW, yielding photon count rates ~0.001-0.01% of the laser repetition rate. Measurement of the cavity spectrum was performed at a laser excitation power of ~ 5 μW. All reported lifetime measurements were performed prior to measuring the fluorescence spectrum.

**Acknowledgements**

The authors acknowledge support from NSF/NSEC under NSF/PHY-06-46094, the use of NSF/NNIN facilities at Harvard University's Center for Nanoscale Systems, and the use of the hpc computer cluster at Harvard.

**Competing Financial Interests**

The authors have no competing financial interests.

# Figure Captions

**Figure 1 | Gap plasmon nanocavity containing coupled emitters. a,** The cavity is comprised of a Ag nanowire and a Ag substrate separated by a gap of thickness $d_G$ set by a dielectric bilayer of $SiO_2$ and the fluorescent dye $Alq_3$. **b,** Cross-section of a cavity showing strong confinement of electric field within the gap. Plotted to the left and bottom are line cuts taken along the vertical and horizontal broken lines, respectively. **c,** SEM micrograph of a cavity (scale bar: 500 nm). Solid circle: approximate diameter and location of incident laser during a representative off-cavity measurement. Broken circle: approximate location during on-cavity measurement. Because the laser spot is larger than the cavity, the on-cavity measurement includes a significant background of off-cavity fluorescence.

**Figure 2 | Spectral and temporal characteristics of cavity-coupled $Alq_3$ fluorescence. a,** Fluorescence spectra of $Alq_3$ on glass slide (Glass); of $Alq_3$ on 12 nm SiO2 on Ag (sample B, Off Cavity); and of $Alq_3$ in a cavity with 15 nm gap (sample B, Cavity). Wavelengths shorter than ~540 nm are blocked by a filter. **b,** Time-resolved fluorescence measurements from the same samples as in (**a**). Points correspond to raw acquired data; solid lines correspond to data corrected for the finite time response of the detector (Supplementary Section 1). Inset: detail of short-time region on linear intensity scale.

**Figure 3 | Cavity enhancement characteristics. a,** Time-integrated intensity of cavity lifetime component of time-resolved fluorescence data from all measured cavities on samples A (5 nm Cav.), B (15 nm Cav.), and C (25 nm Cav.). Broken line indicates dependence expected from cavity enhancement and constant non-radiative lifetime $\tau_{NR} \sim 62$ ps. Solid line indicates dependence expected for loss and constant radiative lifetime. Error bars represent uncertainty in parameter estimation from fits (Supplementary Section 1). **b,** Ratio of time-integrated intensity of cavity component to time-integrated intensities of background components ($I_C/I_B$). **c,** Histogram of total enhancement of peak intensity: peak of cavity lifetime component ($a_C$) normalized to peak intensity from Alq$_3$/glass and multiplied by 6 to approximately account for the difference in number of contributing Alq$_3$ molecules (Supplementary Section 3).

**Figure 1**

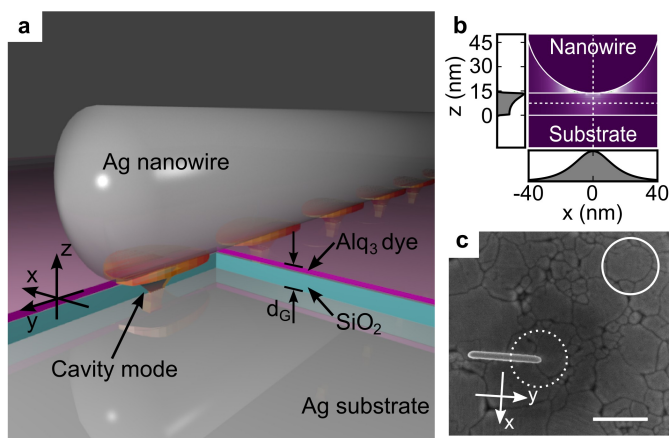

**Figure 2**

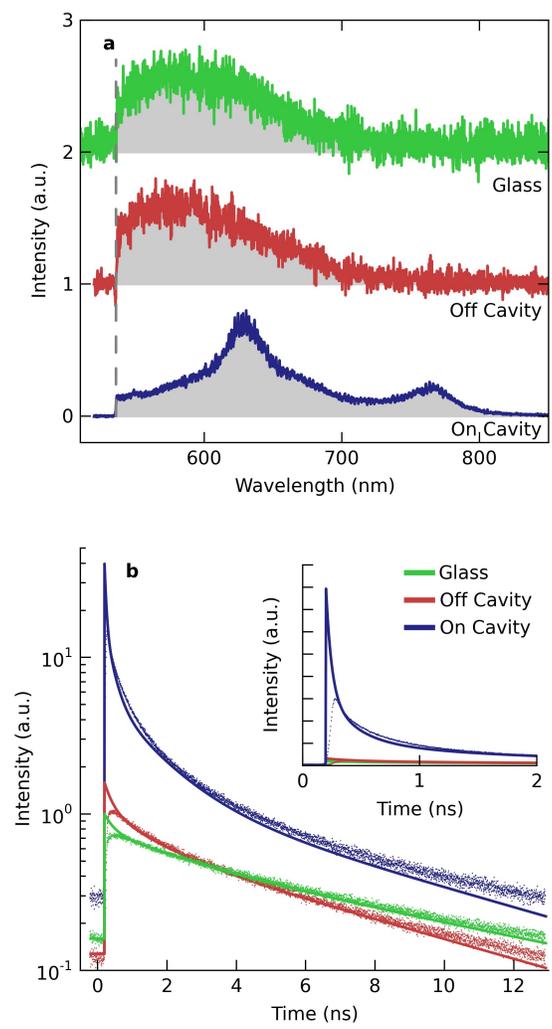

**Figure 3**

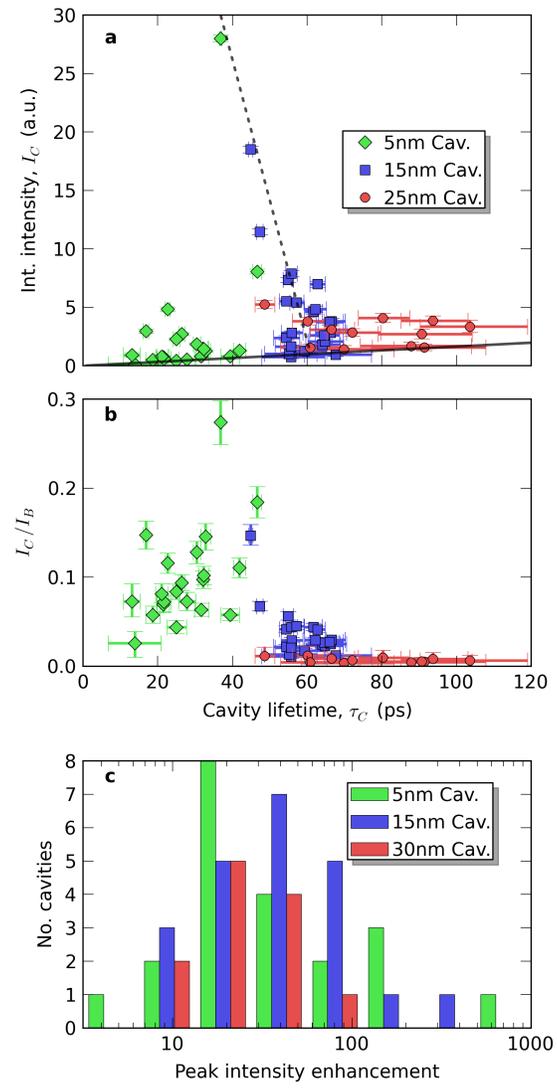

# Supplementary Information

**1. Analysis of time-resolved fluorescence data**

The data measured on cavities displayed a bright, short-lifetime component on top of a background of longer-lifetime components, with the longer-lifetime components having almost identical lifetimes as in data from uncoupled areas of the sample (Fig. S1). Because our measurements of a cavity necessarily include a large background of emitters from the surrounding area that are not coupled to the cavity (Fig. 1c), we concluded that the short-lifetime component of the cavity measurement was due to the cavity and the longer-lifetime background was due to uncoupled emitters. In reality, there is not likely a single cavity lifetime but rather a distribution of cavity lifetimes, but we obtain good fits with our simpler model, so fitting a more complex model is not justified. To the extent that this model is inaccurate, it should tend to reduce the appearance of any cavity effects because any cavity emission that is unable to be described within a single exponential will be fit through an increase in amplitude of one of the background components (the lifetimes of the background components are held fixed during fitting, as described below).

Any measurement of TCSPC fluorescence data is necessarily a convolution of the actual fluorescence signal and the finite time response of the detector[1]. If the fluorescence time scales are much longer than the width of the instrument response function (IRF), then the IRF can be well approximated as a delta-function, and the raw data will provide a good approximation of the actual fluorescence decay. However, in our case we observe measured decay times in the raw data that are not significantly longer than the width of the IRF (Fig. S1a), indicating that the raw data is not a good approximation of the actual fluorescence decay.

Following the conventional approach in fluorescence spectroscopy, we used iterative reconvolution with non-linear least squares fitting to correct the data for the finite width of the IRF[1].

The IRF used for iterative reconvolution was measured using a laser wavelength of ~ 760 nm and an intensity comparable to that of the measured fluorescence signal. Our analysis combined the effects due to the wavelength dependence of the propagation time within the optical fiber used for confocal collection; of the APD response; and of the non-radiative energy relaxation time from the laser excitation energy to the spontaneous emission energy within the emitters. These were treated together using a corrected IRF that was first convolved with a gaussian function having a standard deviation ($\sigma_{IRF}$) chosen to give a good fit to the rise-time of the measured data. We found $\sigma_{IRF}$ ~ 20-25 ps, 10-15 ps, and 0-5 ps, for samples A, B, and C, respectively. This dependence on sample is in qualitative agreement with the observed $Alq_3$ emission spectra from the different samples, which showed broader emission with larger relative intensity of the long-wavelength components for cavities from sample A (Fig. 2a). A representative example of the result of a fit using iterative reconvolution is shown in Fig. S1.

The short cavity lifetime we observe required fine timing resolution (4ps) and high laser repetition rate (76 MHz) to achieve a good signal-to-noise ratio within an integration time short enough to avoid sample drift. This high repetition rate, corresponding to a time of ~13 ns between pulses, prevents the long-lifetime components of the emission from fully decaying. This effect became especially pronounced in the measurement of long-lifetime, uncoupled PbS quantum dots (Fig. S3). We included this effect within our model using a geometric series to account for the fluorescence decay from an infinite series of previous pulses.

The fluorescence dynamics of $Alq_3$ are complicated by loss channels inherent in the evaporated film[2,3]. All time-resolved, off-cavity measurements of $Alq_3$, including measurements on $Alq_3$/glass, required a model containing 3 exponential functions for an adequate fit. This complicated dependence is typical of evaporated films of $Alq_3$ and results from various loss mechanisms such as defects associated with ultra-violet-enhanced photo-oxidation[3]. The radiative lifetime is estimated to be as high as 150 ns, with a typical quantum yield of ~ 0.2[2].

Fits to cavity fluorescence required an additional exponential function to capture the short-time dependence associated with the cavity, increasing the complexity of the model to 4 exponential functions. It would be questionable to fit such a complicated model to our data and expect to obtain useful parameter estimates if all parameters within the model were allowed to vary during the fitting. We observed, however, that on a given sample the lifetimes of the 3 longest-lifetime decays were approximately constant across all measurements, whether the measurement was taken on-cavity or off-cavity (Fig. S1). The first step in the fitting was therefore to find the best-fit values for the 3 long-time decays that minimized the reduced $\chi^2$ simultaneously across all measurements on a sample. These values for the 3 longest lifetimes were then held fixed for fits to individual cavities.

Error bars in Fig. 3 represent conservative estimates of the error in the parameters returned by the fit. These were calculated according to the method outlined in Ref. 1, in which the data set is repeatedly fit with the parameter of interest held fixed at some value near the best-fit value and other parameters free to vary. The error bar is determined by finding the values of the parameter for which the F statistic of the fit exceeds a value determined by the number of free parameters (which in our case is 3282, the number of bins in a TCSPC histogram). This error bar is larger, and probably a better estimate of the actual parameter uncertainty, than the mean squared error estimated using the covariance matrix of the fit[1].

**2. Collection efficiency of measurements of Alq$_3$/glass**

There is an obvious difference in collection efficiency between an off-cavity measurement and a measurement of Alq$_3$ on glass. An off-cavity measurement is effectively performed on a mirror so that all far-field fluorescence is directed upward. On glass, however, light can escape through the substrate.

Using a sample of Alq$_3$ evaporated on a glass cover slip, we measured the amount of light emitted through the substrate. The sample was inverted, and both excitation and collection were performed through the substrate using an oil-immersion objective of numerical aperture 1.25. The intensity was found to be ~2.2 times higher when measured through the glass.

## 3. Simulation of antenna effects

If non-radiative loss was responsible for the short cavity lifetime in our devices, then significant antenna effects would be needed to explain the high cavity intensity we observed. To evaluate this possibility, we performed finite-difference time domain simulations (FDTD Solutions, Lumerical) of absorption as a function of position within the plane corresponding to the Alq$_3$ layer. The cavity simulated had a nanowire length of 900 nm and diameter of 100 nm. Absorption in the Alq$_3$ layer was modeled with a 3 nm-thick layer of GaAs on top of an insulating, non-absorbing dielectric layer (representing the SiO$_2$ layer of the actual cavities). Optical constants for the Ag substrate and nanowire were taken from from Johnson & Christy[4]. The excitation source was modeled as a broadband Gaussian pulse focused onto the end of the nanowire, as in Fig. 1c.

The resulting absorption was integrated according to distance from the nanowire: that occurring within $d_G$ of the nanowire was considered "under" the nanowire, whereas everything farther than $d_G$ was considered background. These simulations were repeated for all four sample structures, with two simulations on the sample C structure: one with and one without a nanowire. For the simulations of glass substrate and of an off-cavity region of sample C, the determination of absorption "under" the nanowire was made by integrating over an area of the film of the same size and position as in the simulation of sample C that included a nanowire. The absorption in this area is seen to be ~1/6 of the total absorption, which indicates that, since our excitation is far below saturation, there are approximately 1/6 as many molecules excited in the area under the nanowire than in the full laser spot.

Therefore, to find the total intensity enhancement (Fig. 3c), we multiplied the peak intensity measured on glass by 1/6.

The results of the simulations make apparent that the nanowire structure does not lead to significantly enhanced absorption in the region under the nanowire (Fig. S2). In fact, the simulated absorption under the nanowire is highest for the sample that experimentally gave the lowest cavity emission (sample C). Most importantly, the simulations of sample C on and off-cavity show that the nanowire leads to significantly more background absorption but less absorption under the nanowire. We can therefore conclude that antenna effects from the nanowire cannot explain the high intensity of the measured cavity emission.

The simulations also suggest that polarization of the excitation laser should affect the measured cavity intensity, but we found no dependence of cavity intensity with cavity orientation relative to the (fixed) laser polarization. The intensity variations that we observe are larger than the polarization effects predicted by simulation (Fig. 3a), suggesting that the observed differences in intensity are due to other differences that we are unable to adequately probe, such as the nanoscale geometry of the end of the nanowire or the uniformity of the gap spacing.

**4. Measurements of cavities with PbS quantum dot emitters**

To further illustrate the wide applicability of our cavity design, we performed measurements on samples utilizing layers of PbS colloidal quantum dots (Evident, ~850 nm center wavelength, ~200 nm spectral width) instead of Alq$_3$ (Fig. S3). Ag templates coated with 2 nm SiO$_2$ were prepared as for the Alq$_3$ samples, but instead of evaporating Alq$_3$, PbS nanocrystals were deposited by spin-coating at 3000 r.p.m. for 45 s from a droplet of toluene containing 200 μM concentration of PbS nanocrystals. Nanowires were then transferred on top of the layer of nanocrystals as for the other samples. These samples show a nearly mono-exponential decay with a very long lifetime (~80 ns) when not coupled to a cavity. As a result, the short-lifetime cavity decay becomes especially pronounced. Iterative reconvoution fits using a three-exponential model find a typical uncoupled lifetime of ~80 ns, a typical cavity lifetime of ~0.04 ns, and thus an emission rate enhancement of ~2000.

Off-cavity measurements were performed in an open area of the sample ~ 5μm away from the cavity, ensuring approximately the same density of PbS quantum dots in the regions of both measurements.

While the decay dynamics are less complicated in these samples than in the Alq$_3$ samples, the thickness of the PbS quantum dot layer is difficult to control using spin-coating, and it is likely to be non-uniform. This prevents us from being able to perform a meaningful comparison between samples of different gap spacings (or even simply knowing the gap spacing), as is possible with the Alq$_3$ samples. We have previously fabricated samples with uniform layers of PbS quantum dots using a stamping method[5], but this approach was found to damage the quantum dot layer, creating short-lifetime decay pathways that made the quantum dots unsuitable to characterize the cavity lifetime.

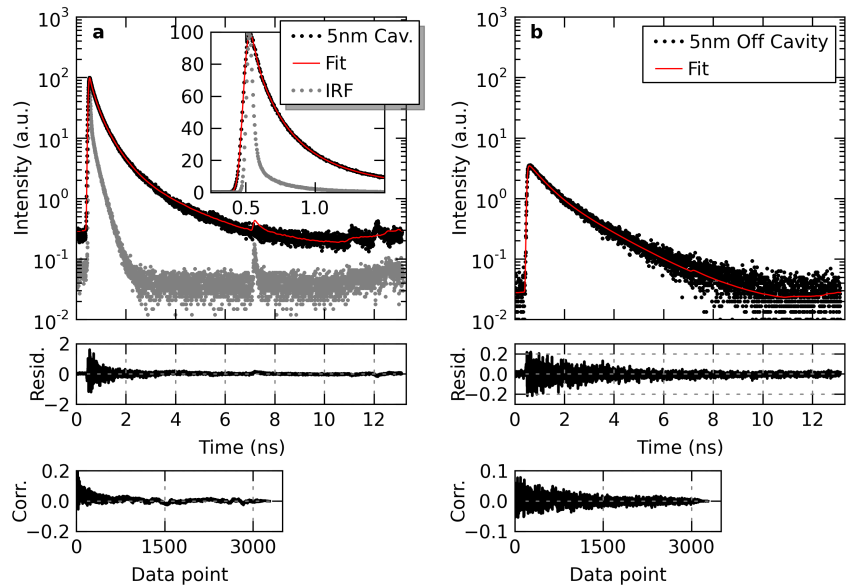

**Figure S1 | Iterative reconvolution of exponential decay model fitted to time-resolved fluorescence data from sample A. a,** Fit to data from a cavity (5nm Cav.). The lifetimes of the 3 longest-lifetime components were determined using a global fit to all measurements from sample A (including off-cavity areas) and were used as fixed parameters for the fit shown here. Upper panel: Data and reconvolved, fitted model (Fit). Also shown for comparison is the measured IRF (IRF). The small peaks occurring at long times are due to reflections within the optical fiber used for confocal collection. Inset: detail of short-time region. Middle panel: Residuals of the fit (difference between data and Fit). Lower panel: Autocorrelation of the residuals. **b,** Fit to off-cavity data from sample A (5nm Off cavity) using the same long-lifetime components as for the fit in (**a**). Here all three lifetimes were held fixed during fitting, allowing only the amplitude coefficients to vary.

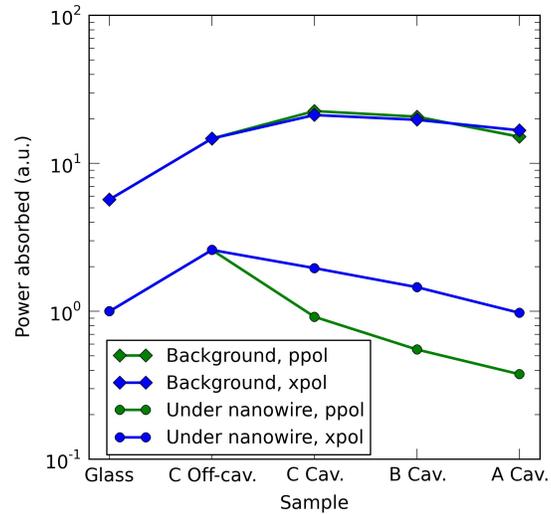

**Figure S2 | FDTD simulations of absorption of a gaussian pulse of 448 nm wavelength positioned over the end of a 100 nm-diameter nanowire.** Simulations were repeated for laser polarization parallel to (ppol) and perpendicular to (xpol) the axis of the nanowire. The total absorption has been separated into two components: that "under" the nanowire and that of the background (i.e. total minus that under the nanowire). The absorbing layer is approximated as a 3nm-thick layer of GaAs. For sample C, results are shown from simulations both with and without a nanowire (C Cav. and C Off-cav., respectively). For simulations lacking a nanowire (Glass and C Off-cav.), the absorption was integrated over the same area as the nanowire cavity in sample C.

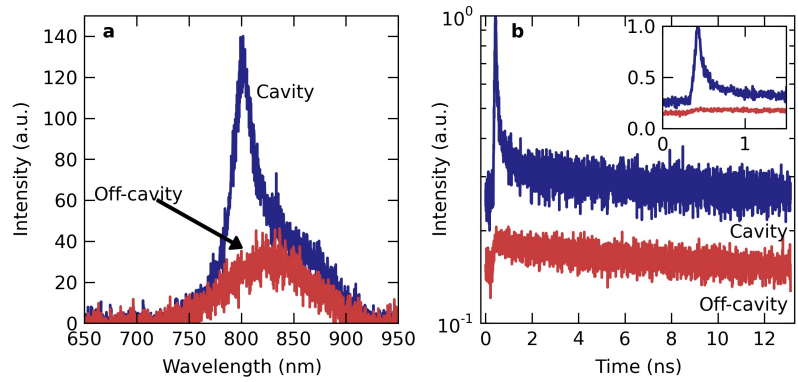

**Figure S3 | Spontaneous emission enhancement in a cavity containing PbS colloidal quantum dots in place of Alq$_3$. a,** Spectra of PbS collected at 800 nW (Cavity) and 6μW (Off-cavity). **b,** Time-resolved fluorescence measurements acquired at 330 nW. Both Cavity and Off-cavity data are normalized to the peak of the cavity trace. Inset: Detail of short-time behavior on linear vertical scale.

**Supplementary References**